\documentclass[12pt, preprint]{aastex}

\begin{document} 

\title{GW Librae: Still Hot Eight Years Post-Outburst}

\author{Paula Szkody\altaffilmark{1,2},
Anjum S. Mukadam\altaffilmark{1,2},
Boris T. G\"ansicke\altaffilmark{3},
Paul Chote\altaffilmark{3},
Peter Nelson\altaffilmark{4},
Gordon Myers\altaffilmark{4},
Odette Toloza\altaffilmark{3},
Elizabeth O. Waagen\altaffilmark{4},
Edward M. Sion\altaffilmark{5},
Denis J. Sullivan\altaffilmark{6},
Dean M. Townsley\altaffilmark{7}}

\altaffiltext{1}{Department of Astronomy, University of Washington,
  Box 351580, Seattle, WA 98195; szkody@astro.washington.edu}
\altaffiltext{2}{Based on observations obtained with the Apache Point
  Observatory (APO) 3.5-meter telescope, which is owned and operated
  by the Astrophysical Research Consortium (ARC).}
\altaffiltext{3}{Department of Physics, University of Warwick, Coventry CV4 7AL, UK}
\altaffiltext{4}{AAVSO, 48 Bay State Rd, Cambridge MA 02138}
\altaffiltext{5}{Department of Astrophysics and Planetary Science, Villanova University, Villanova, PA 19085}
\altaffiltext{6}{School of Chemical \& Physical Sciences, Victoria University of Wellington, P. O. Box 600, Wellington, NZ}
\altaffiltext{7}{Department of Physics and Astronomy, University of Alabama, Tuscaloosa, AL 35487}
 
\begin{abstract}

We report continued Hubble Space Telescope (HST) ultraviolet spectra and ground-based optical
photometry and spectroscopy of GW Librae eight years after its largest known dwarf
nova outburst in 2007. This represents the longest cooling timescale measured for any
dwarf nova. The spectra reveal that the white dwarf still remains about 3000 K
hotter than its quiescent value. Both ultraviolet and optical light curves show a
short period of 364-373 s, similar to one of the non-radial pulsation periods present for 
years prior to the outburst, and with a similar large UV/optical amplitude ratio. 
A large modulation at a period of 2 h (also similar to that observed prior
to outburst) is present in the optical data
preceding and during the HST observations, but the satellite observation intervals did
not cover the peaks of the optical modulation so it is not possible to determine its
corresponding UV amplitude. The similarity of the short and long periods to quiescent
values implies the pulsating, fast spinning white dwarf in GW Lib may finally be nearing its 
quiescent configuration.
\end{abstract}

\section{Introduction}

GW Librae was known as an ordinary low accretion rate dwarf nova with infrequent 
large amplitude outbursts (Gonzalez \& Maza 1983) and a very
short orbital period of 76.78 min (Thorstensen et al. 2002) until it
became highlighted as the first accreting white dwarf in a cataclysmic
variable to show non-radial pulsations (Warner \& van Zyl 1998). Further
monitoring at quiescence over several years revealed relatively stable 
pulsations at 648, 376 and 236 s (van Zyl et al. 2000, 2004) and Hubble Space Telescope (HST)
ultraviolet observations showed the same periods with higher amplitudes (Szkody
et al. 2002), consistent with the source of the variation being modulation of the
temperature of the white dwarf photosphere (Robinson et al. 1995).
The temperature of the white dwarf in GW Lib at quiescence was determined
from the HST spectra to be near 15,000 K (for log g=8). Although this temperature is
 outside the normal instability 
strip for ZZ Ceti pulsators with a pure hydrogen atmosphere, it is within the instability
strip(s) for accreting white dwarf pulsators that have an atmosphere with a solar
composition (Arras et al. 2006).

In 2007 April, GW Lib underwent a second outburst of 9 mag (Templeton et al.
2007), the largest known for any dwarf nova. Subsequent optical and
ultraviolet observations have provided a long term record of the impact
of this large outburst amplitude on the white dwarf. The heating/cooling
and its effect on the white dwarf pulsations have now been followed for
8 years. Ground based optical observations were available over most 
years (Copperwheat et al. 2009, Schwieterman et al. 2010, Bullock et al. 2011,
Vican et al. 2011, Szkody et al. 2012, Chote \& Sullivan 2016), and the
ultraviolet monitoring took place with $\it{GALEX}$ in 2007-2010 (Bullock et al. 2011)
and with HST in 2010, 2011 (Szkody et al. 2012) and 2013 (Toloza et al. 2016).
Both of these wavelength regions showed some interesting and surprising results.

In the optical, a period of 296 s was marginally detected on one night in 2008 June  
(Copperwheat et al. 2009) at the 10 millimodulation amplitude (mma) level. The next time this 
short period (280-290 s)
was seen was in optical data obtained in 2010 March, 2011 April, and 2012 May
(Szkody et al. 2012, Chote \& Sullivan 2016) with amplitudes of 9 mma and in HST ultraviolet 
data in 2010 March and 2011 April with amplitudes of 20 and 50 mma respectively. 
Strong signals (25 mma) at 19 minutes
were evident in optical data taken by several groups throughout 2008 Mar-July but then this
period disappeared until reappearance in 2012 Apr-June (Chote \& Sullivan 2016) at the 
50 mma level. An even longer period at about 2 hr was identified
prior to outburst (Woudt \& Warner 2002, Copperwheat et al 2009, Hilton et al. 
2007), and strong modulations at periods of 3-4 hr were observed after outburst in both optical
and $\it{GALEX}$ UV data (Schwieterman 
et al. 2010, Bullock et al. 2011, Vican et al. 2011, Chote \& Sullivan 2016, Toloza 
et al. 2016).

The temperature of the white dwarf as determined from HST spectra are a
function of the gravity (mass) assumed. Recently, Toloza et al. (2016)
reanalyzed all the available HST spectra of GW Lib using a common log g=8.35, 
which is consistent
with the most recent mass estimates of 0.8 M$_{\odot}$ (van Spaandonk et al. 2010, Szkody 
et al. 2012). They obtained values of 14,695 K from the 2002 quiescent data,
and 17,980 K in 2010, and 15,915 K in 2011 for the 3 and 4 year post-outburst data. Each of 
these values showed a
500 K variation in temperature in spectra phased at the peak versus the troughs of the 
short period pulsations that were present.
Surprisingly, the three orbits of HST data in 2013 showed much larger
changes in flux (a factor of 2) and temperature (15,975-18,966 K), with a 
mean temperature (16937 K) larger than the 2011 data. The
large flux changes appeared to be related to the 4 hr variability that was 
evident at that time.

In order to continue to monitor GW Lib during its return to quiescence,
we obtained further HST and optical observations in 2015.

\section{Observations}

Once the HST date was set, ground-based observations were coordinated 
with nights before and during the observation. The observations
obtained are summarized in Table 1.

\subsection{HST Data}

Three HST orbits on April 21 were used to collect data with the Cosmic Origins
Spectrograph (COS) using the G140L grating in time-tag mode. Useful spectra
were obtained from 1130-2020\AA, with a resolution of about 0.75\AA.
Light curves were created by summing the fluxes over all the continuum
wavelengths in this range in 5 s bins, leaving out the strong geocoronal 
emission line of Ly$\alpha$ and the strong CIV emission line from GW Lib.
These light curves were then divided by the mean and one was subtracted so
that a fractional amplitude scale was produced that could be used for
Discrete Fourier Transform (DFT) period analysis. The amount of noise was
determined by a shuffling technique to find a 3$\sigma$ limit (see
Szkody et al. 2012 for further details).

\subsection{Optical Data}

The American Association of Variable Star Obervers (AAVSO) posted alerts
and monitored the optical brightness prior to the HST observations to
ensure the system remained at quiescence. The mean magnitude during 
April was 16.7. Optical photometry was accomplished on Apr 21 and 22
using the 3.5 m telescope at Apache Point Observatory (APO) and the 1 m
telescope at the University of Canterbury Mt.\ John Observatory (UCMJO). Instruments at
both places incorporated similar frame transfer CCDs with negligible time
lost to readout, and a BG-40 broad band blue filter: Agile at APO 
(Mukadam et al. 2011) and Puoko-nui at UCMJO (Chote et al. 2014). Cloudy
weather resulted in lower quality data at APO on Apr 22 compared with Apr 21.

The APO optical reductions were
accomplished using standard IRAF\footnote{IRAF is distributed by the National
Optical Astronomy Observatory,
which is operated by the Association of Universities for Research in 
Astronomy, under cooperative agreement with the National Science 
Foundation.} routines to extract sky-subtracted light curves from the CCD frames
using weighted circular aperture photometry (O'Donoghue et al. 2000). 
For the short period analysis of the APO data, the light curves were
converted to fractional amplitude in the same manner as for the HST data.
The UCMJO data utilized the reduction pipeline tsreduce described in Chote et al. (2014).
 
Spectra were obtained on 2015 April 21 using the Double Imaging Spectrograph 
(DIS) at APO.
The high resolution grating was used to provide
simultaneous blue and red spectral coverage with a resolution of
0.6\AA\ pixel$^{-1}$ for blue
wavelengths of 4000-5000\AA\ and red wavelengths of 6000-7200\AA. 
Flux standards and HeNeAr lamps were used for calibration and
 the IRAF tasks
under ${\it ccdproc, apall}$ and ${onedspec}$ were used to correct
the images, extract the spectra to 1-d and calibrate them. 

\section{Results}

\subsection{HST Ultraviolet and APO Optical Spectra}

Figure 1 shows the average spectrum from the three HST orbits in 2015 overplotted on the
average of the three orbits from 2013. These average spectra separated by 
two years are very similar. Using the same procedure to fit the average
spectrum in 2015 as done for all the previous data (Toloza et al. 2016) results in a mean temperature of 17560$\pm$9 K. This implies that the white dwarf has 
still not returned to quiescence eight years after its outburst. The optical 
spectrum obtained
24 hrs prior to the HST spectra is shown in Figure 2. The overall spectral 
shape is similar to the quiescent spectra taken with the same spectrograph 
(Szkody, Desai \& Hoard 2000)
while the blue fluxes are between the quiescent values and those obtained in 2010
(Szkody et al. 2012). The FWZI of H$\beta$ (20\AA) is wider while the equivalent width
(18\AA) is smaller than quiescent values, numbers that are consistent with a higher
temperature white dwarf and a larger contribution from the inner higher velocity disk
regions.

\subsection{Optical and UV Light Curves}

The optical light curves from APO, UCMJO and the AAVSO, as well as the UV light
curve constructed from the HST spectra are shown in Figure 3. Constants were
added to the magnitudes of the APO, UCMJO and HST light curves to bring them all to the 
approximate AAVSO magnitude for each night.
The optical data show a consistent 20\% amplitude modulation at 2 hr that 
persists from the preceding night through the time of the HST observations. 
While the UV shows a mean change of about 10\% over the 3 orbits, the times 
of peak optical flux unfortunately did not fall into the HST observation 
windows. Thus, it is impossible to tell if the large increase seen in the 
2013 UV data existed in 2015. However, the length of the large UV flux 
increase in 2013 was at least 100 min  and the gaps in the 2015 data are
only about 50 min so some increase should have been visible if the same 
phenomenon was present. The optical and especially the UV light 
curves do show the presence of a shorter timescale variation.

\subsection{Optical and UV Pulsations}

The DFT results are shown in Figure 4 for the
UV data and Figures 5 and 6 for the APO data obtained on  Apr 21 and simultaneous with HST 
on Apr 22. All datasets show a significant period between 364-373 s, one of the periods
that is visible before the 2007 outburst. The UV/optical amplitude ratio is 100/15 = 6.7, a
similar ratio to that observed at quiescence for the 376 s period (Szkody et al. 2002).
  
\section{Discussion}

Prior studies have addressed the question of the origin of the three main periodicities
visible after the outburst. While the short 280-370 s periods are usually ascribed
to a non-radial pulsation mode, the origin of the intermittent longer periods at 19 min 
and 2-4 hr have been harder to interpret as due to pulsations or quasi-periodic 
oscillations of
the accretion disk. The Chote \& Sullivan (2016) observations obtained 
over a timescale of 3 months in 2012 and 
their interpretation 
support a pulsation mode for the 19 min period. Their arguments hinge on the similarity
of the period between 2008 and its return in 2012, and the similarity of the behavior
of the period (amplitude modulation and slight frequency shifts) to that seen
in cool DAV stars (Kleinman et al. 1998) and to the flare events that repeat every few
days in DAV white dwarfs recently reported by Bell et al. (2015) and Hermes et al. (2015).

Toloza et al. (2016) also argue that the 2-4 hr modulation that appears and 
disappears is related to pulsations. They fit the  large amplitude of the variation 
with an increase in the temperature of the white dwarf over a fraction of 
the white dwarf surface. They speculate this variation could be caused by a splitting of 
the g-modes due to
the rapid rotation of the white dwarf in GW Lib (200 s, Szkody et al. 2012), that results
in a travelling wave moving counter to the rotation. In both cases, the similarity of the
periods when they are present, yet the lack of a regular recurrence time rule out
phenomena such as disk precession or beating between periods.  

However, the theoretical 
details of the long period pulsations remain to be delineated. These include
reasons why the 19 min period remains for months and then disappears for years, why the
longest period changes from 2 hr (Woudt \& Warner 2002, this work) to 3 hr (Chote \& Sullivan
2016) to 4 hr (Bullock et al. 2011, Toloza et al. 2016), and why the 4 hr optical variation
can disappear from one night to the next and be out of phase with the ultraviolet (Bullock
et al. 2011). It is possible the changes in period may be related to the outburst and
subsequent cooling.
The period of 2 hr was evident prior to outburst, then it was 4 hr in 2008-2010, 3 hr in 
2012 and 2 hr in 2015. 
 
\section{Conclusions}

Our HST ultraviolet and ground-based optical coverage of GW Lib 8 years after
the largest known dwarf nova outburst reveals that the white dwarf has not
yet reached its quiescent pre-outburst temperature. Its mean temperature for
the 2015 observation remains similar to what it was in 2013, about 3000 K above
its quiescent value. The ultraviolet and optical light curves both show a
short period of 364-373 s, similar to one of the persistent periods observed during
quiescence, and with a similar ratio (7) of UV/optical amplitudes. A large
(0.2 mag peak-to-peak) modulation at a period of 2 hr is apparent in the optical light 
curves preceding and simultaneous with the HST data, and is coherent over these 2 nights. 
Unfortunately, the HST observation times did not cover the peaks of the optical
modulation so it is not possible to tell if, or how, the 2 hr modulation appears 
in the UV. Neither the
19 min period that was evident in the optical in 2012 nor the large 4 hr modulation that
was present in the 
2013 HST data are observed. The 19 min period has yet to be seen at ultraviolet wavlengths
and a much longer series of optical and ultraviolet observations will be needed to sort out
the recurrence timescales and wavelength dependence of the 2-4 hr modulations. The
return of the short and long periods to their pre-outburst values may be a signal
that the white dwarf is finally returning to its quiescent configuration.

\acknowledgments
PS and ASM acknowledge support from NASA grant HST-GO13807 from the Space
Telescope Science Institute, which is operated by the Association of
Universities for Research in Astronomy, Inc., for NASA, under contract
NAS 5-26555, and from NSF grant AST-1514737. We especially thank AAVSO
observers Josch Hambsch, Damien Lemay, and Gary Walker 
for their monitoring of GW Lib.
The research leading to these results has also received funding from the
European Research Council under the European Union's Seventh Framework
Programme (FP/2007-2013) / ERC Grant Agreement n. 320964 (WDTracer).

\clearpage
\begin{deluxetable}{lcccc}
\tabletypesize{\scriptsize}
\tablewidth{0pt}
\tablecaption{Summary of April 2015 Observations}
\tablehead{
\colhead{UT Date} & \colhead{Obs} & \colhead{Instr.} & \colhead{Time} & 
 \colhead{Exp (s)}}
\startdata
21 & APO & DIS & 06:42-07:03 & 2x600 \\
21 & APO & Agile & 07:42:26-11:42:26 & 20 \\
21 & UCMJO & Puoko-nui & 12:59:30-13:32:00 & 30 \\
21 & AAVSO & CCD & 11:52:49-16:15:05 & 120 \\
22 & HST & COS & 05:02:20-05:38:20 & time-tag \\
22 & HST & COS & 06:31:40-07:16:50 & time-tag \\
22 & HST & COS & 08:07:10-08:52:20 & time-tag \\
22 & APO & Agile & 06:58:28-08:49:11 & 20 \\
22 & UCMJO & Puoko-nui & 09:16:30-10:45:00 & 30 \\
\enddata
\end{deluxetable}

\clearpage

\begin{figure}
\figurenum {1}
\plotone{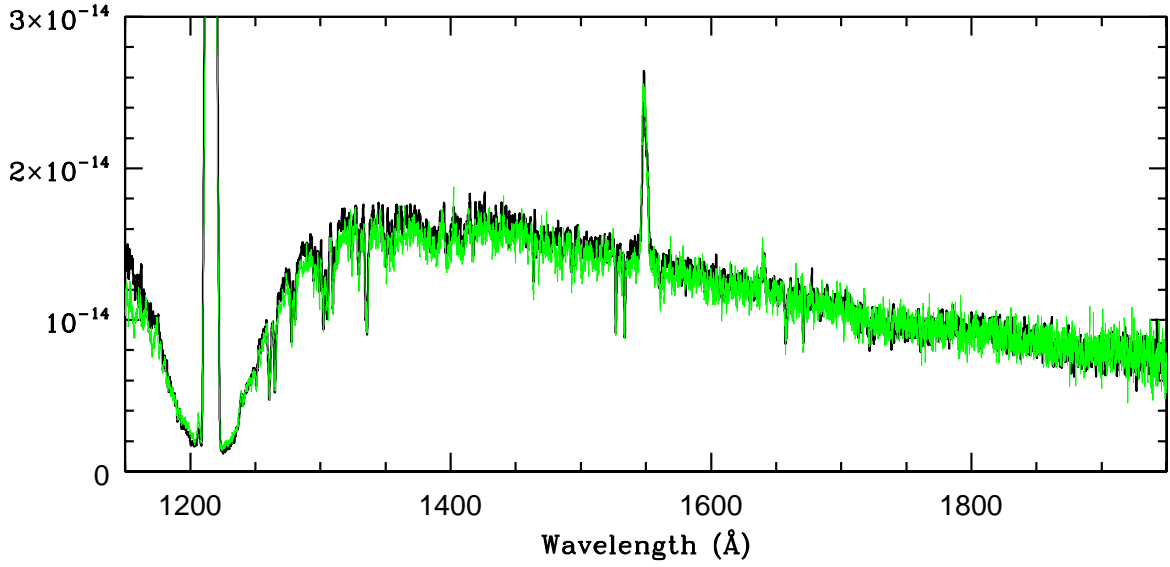}
\caption{The average 2015 April COS spectra from 3 orbits (black) overplotted on the
average 2013 May COS spectra from 3 orbits (green).
Vertical axis is F$_{\lambda}$ in units of ergs cm$^{-2}$ s$^{-1}$ \AA$^{-1}$.}
\end{figure}

\clearpage
\begin{figure}
\figurenum {2}
\plotone{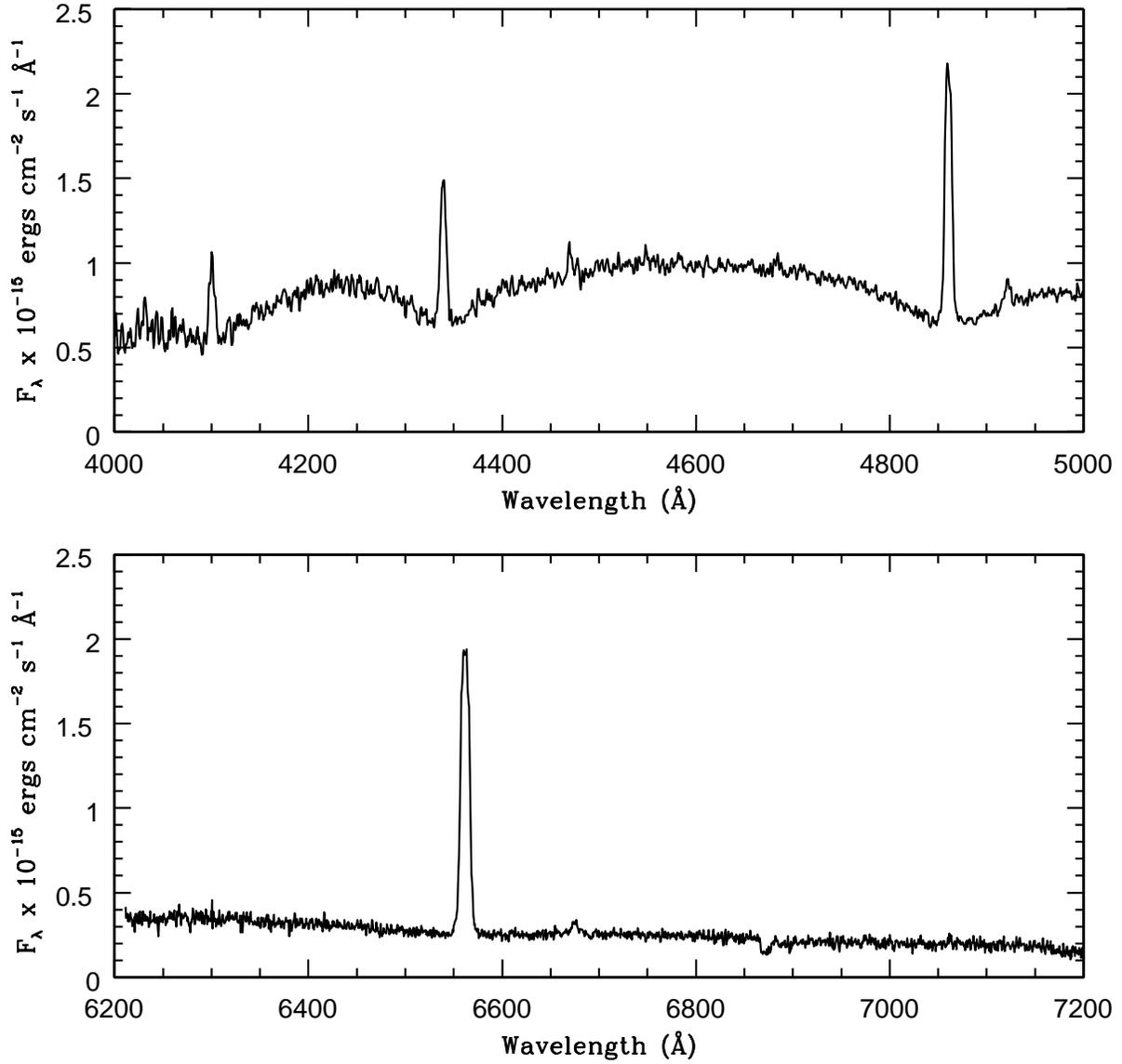}
\caption{DIS blue and red spectra obtained 21 Apr showing the typical Balmer 
emission lines flanked by absorption from the white dwarf.}
\end{figure}

\clearpage
\begin{figure}
\figurenum {3}
\plotone{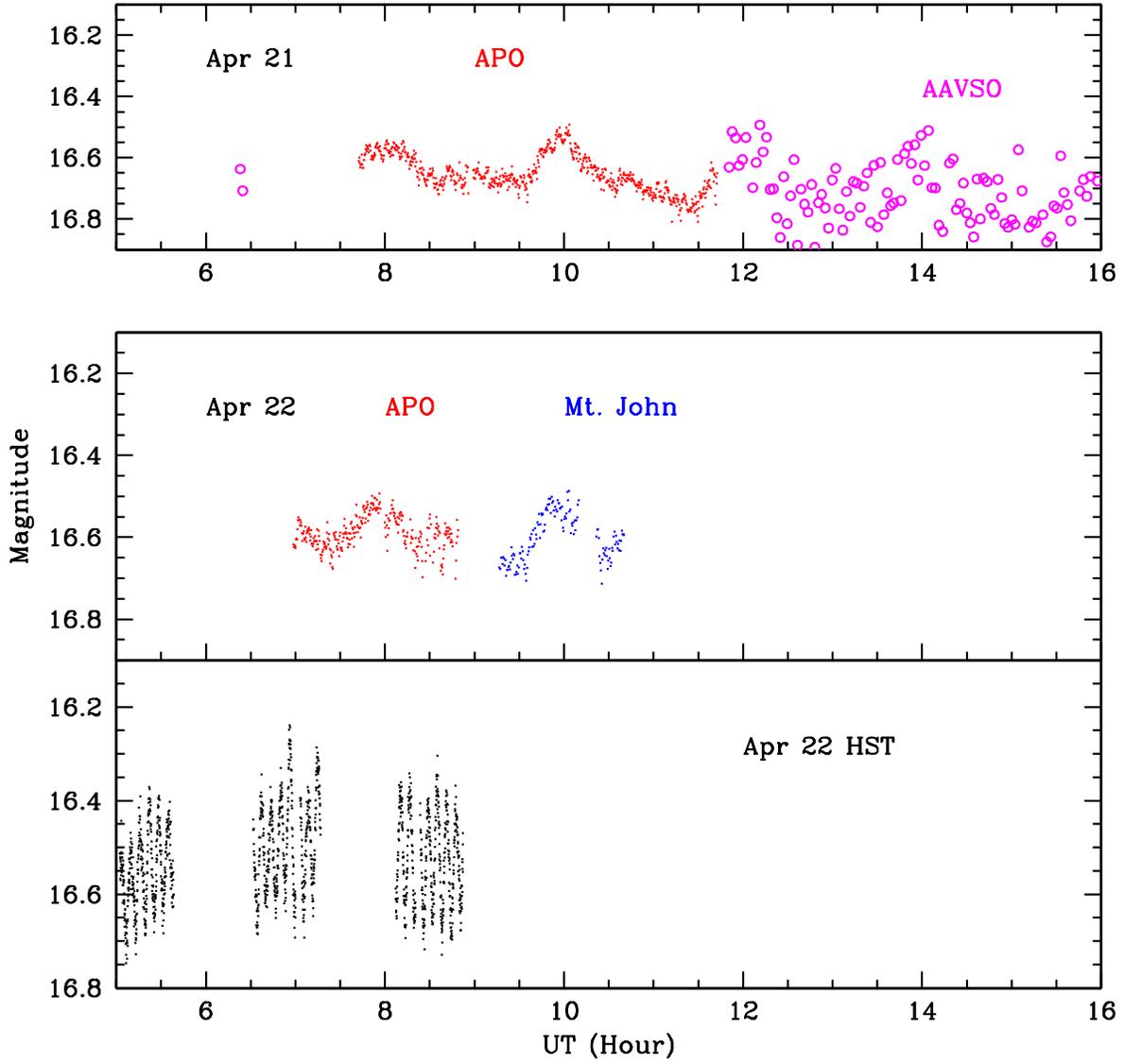}
\caption{Light curves in the optical and UV for Apr 21 and 22nd. The dense
red points are from APO, the open magenta points are AAVSO data, the blue are
 UCMJO and the black are HST.}
\end{figure}

\clearpage
\begin{figure}
\figurenum {4}
\plotone{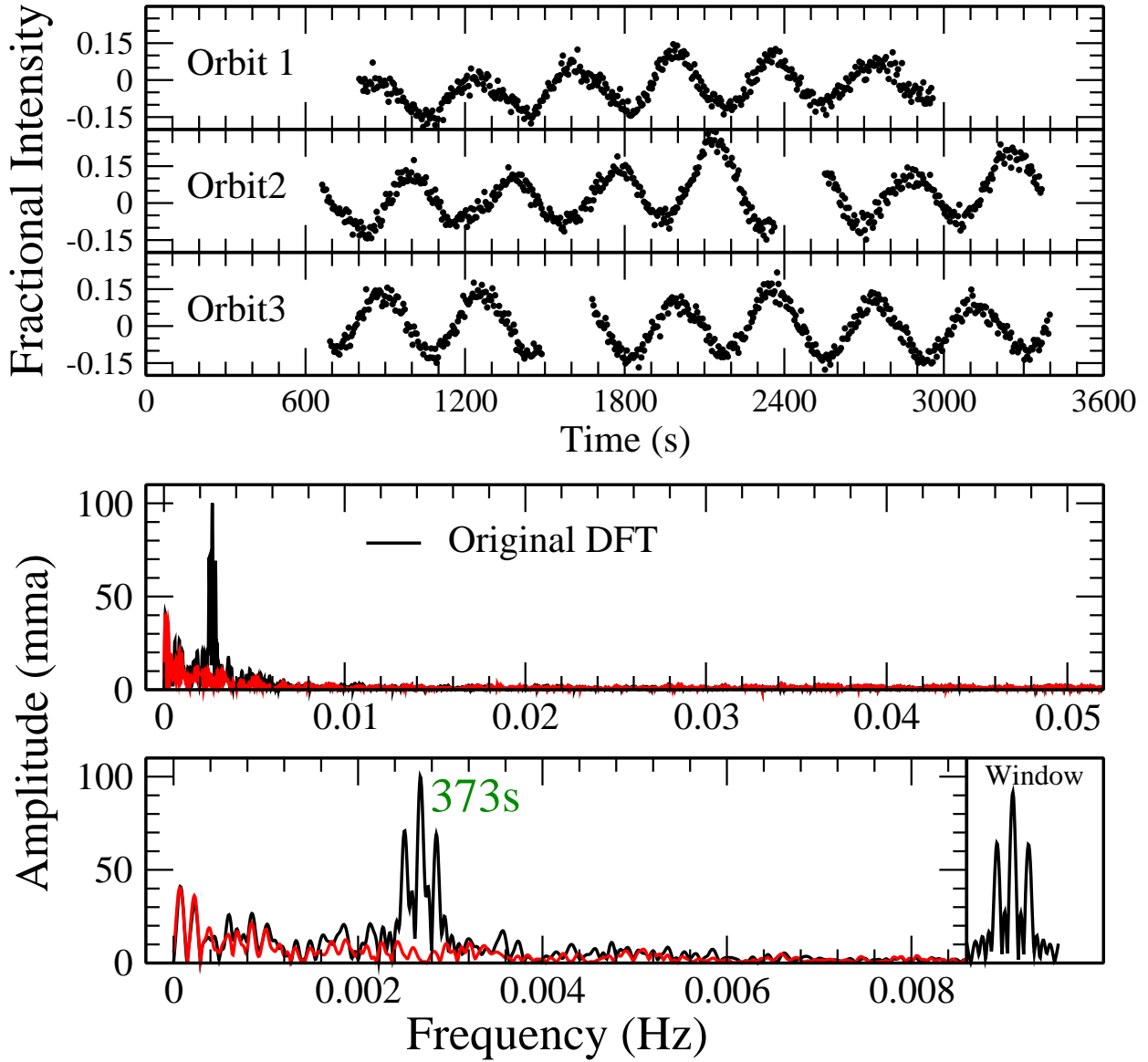}
\caption{Intensity light curves (top) and DFT for the UV data from 
three HST orbits on
22 Apr. Bottom is an expanded area around the main period.}
\end{figure}

\clearpage
\begin{figure}
\figurenum {5}
\plotone{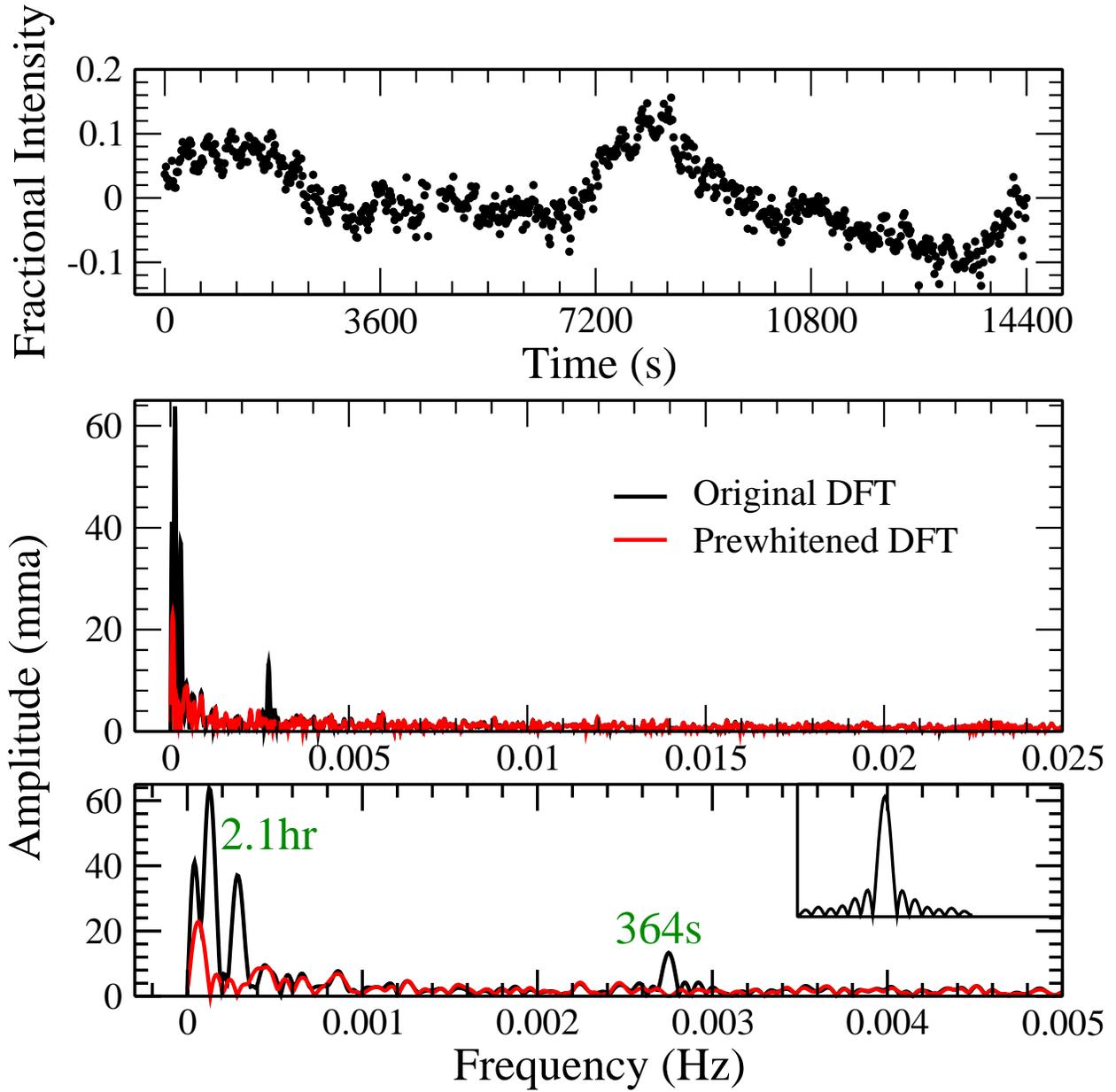}
\caption{Intensity light curve (top) and DFT for the Agile optical data taken
on 21 April.}
\end{figure}

\clearpage
\begin{figure}
\figurenum {6}
\plotone{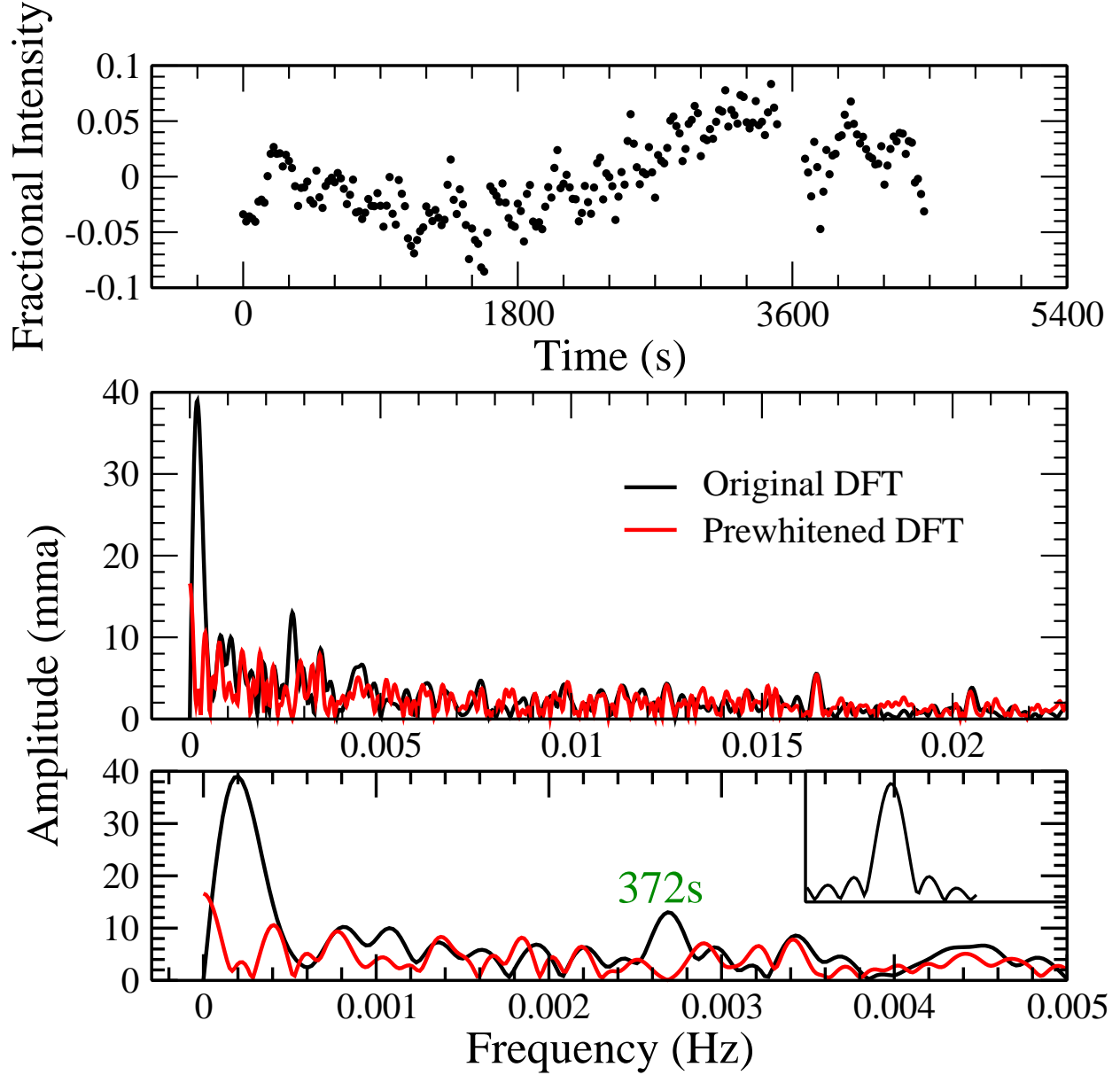}
\caption{Intensity light curve and DFT for the Agile optical data obtained simultaneously with HST.}
\end{figure}
 

\begin{thebibliography}{}

\bibitem[Arras et al. (2006)]{A06} Arras, P., Townsley, D. M., Bildsten, L., 2006, \apj, 643, L119

\bibitem[Bell et al. (2015)]{B15} Bell, K. J. et al. 2015, \apj, 809, 14

\bibitem[Bullock et al. (2011)]{B11} Bullock, E. et al. 2011, \aj, 141,84

\bibitem[Chote et al. (2014)]{C14} Chote, P. et al. 2014, \mnras, 440, 1490

\bibitem[Chote \& Sullivan (2016)]{CS16} Chote, P. \& Sullivan, D. J. 2016, \mnras, 458, 1393

\bibitem[Copperwheat et al. (2009)]{C09} Copperwheat, C. M. et al. 2009, \mnras, 393, 157

\bibitem[Gonzalez \& Maza (1983)]{GM83} Gonzalez, L. E. \& Maza, J. 1983, IAU Circ. 3854

\bibitem[Hermes et al. (2015)]{H15} Hermes, J. J. et al. 2015, \apj, 810, L5

\bibitem[Hilton et al. (2007)]{H07} Hilton, E. et al. 2007, \aj, 134, 1503

\bibitem[Kleinman et al. (1998)]{K98} Kleinman, S. J. et al. 1998, \apj, 495, 424

\bibitem[Mukadam et al. (2011)]{M11} Mukadam, A. S. et al. 2011, \pasp, 123, 1423

\bibitem[O'Donoghue et al. (2000)]{OD00} O'Donoghue, D. et al. 2000, Baltic Astronomy, 9, 375

\bibitem[Robinson et al. (1995)]{R95} Robinson, E. L. et al. 1995, \apj, 438, 908

\bibitem[Schwieterman et al. (2010)]{S10} Schwieterman, E. W. et al. 2010, JSAARA, 3, 6

\bibitem[Szkody et al. (2000)]{S00} Szkody, P., Desai, V. and Hoard, D. W. 2000, \aj, 119, 365

\bibitem[Szkody et al. (2002)]{S02} Szkody, P., G\"ansicke, B. T., Howell, S. B. and Sion, E. M., 2002, \apj, 575, L79

\bibitem[Szkody et al. (2012)]{S12} Szkody, P. et al. 2012, \apj, 753, 158

\bibitem[Templeton et al. (2007)]{T07} Templeton, M., Stubbings, R., Waagen, E. O. et al. 2007, CBET, 922, 1

\bibitem[Thorstensen et al. (2002)]{T02} Thorstensen, J. R., Patterson, J., Kemp, J. and Vennes, S. 2002, \pasp, 114, 1108

\bibitem[Toloza et al. (2016)]{T16} Toloza, O. et al. 2016, \mnras, in press.

\bibitem[Warner \& van Zyl (1998)]{WZ98} Warner, B. \& van Zyl, L. 1998 in IAU Symp. 185, New Eyes to See Inside the Sun and Stars, ed. F.-L. Deubner, J. Christensen-Dalsgaard, \& D. Kurtz (Cambridge:Cambridge Univ. Press), 321

\bibitem[Woudt \& Warner (2002)]{WW02} Woudt, P. A. \& Warner, B. 2002, \apss, 282, 433

\bibitem[van Spaandonk et al. (2010)]{VS10} van Spaandonk, L., Steeghs, D., Marsh, T. R., Parsons, S. G. 2010, \apj, 715, L109

\bibitem[van Zyl et al. (2000)]{VZ00} van Zyl, L. et al. 2000, Balt. Astron., 9, 231

\bibitem[van Zyl et al. (2004)]{VZ04} van Zyl, L. et al. 2004, \mnras, 350, 307

\bibitem[Vican et al. (2011)]{V11} Vican, L. et al. 2011, \pasp, 123, 1156

\end{thebibliography}
\end{document}